\documentclass[aps, prd,nofootinbib,superscriptaddress,reprint,onecolumn,notitlepage,11pt,twoside]{revtex4-1}
\usepackage[ includeheadfoot,paperheight = 269mm,paperwidth = 190mm, top=18mm, left = 28mm, right=18mm, bottom=22mm]{geometry} 
\usepackage{graphicx}
\usepackage{amssymb}
\usepackage{amsmath}
\usepackage{fancyhdr}
\usepackage{amsmath,graphicx,color,mathrsfs}
\usepackage{latexsym}
\usepackage{amsmath}
\usepackage{amssymb}
\usepackage{graphics}
\usepackage{epsfig}
\usepackage{graphicx}
\usepackage{subfigure}
\usepackage{bm}
\usepackage{hyperref}
\usepackage[makeroom]{cancel} 
\usepackage{multirow}
\usepackage{enumitem}

\def\Hinf{H_{\rm inf}}

\def\Ninf{N_{\rm inf}}
\def\kp{k_{\rm pivot}}

\def\omr{\Omega_{\rm r}}
\def\oml{\Omega_{\Lambda}}
\def\omk{\Omega_{k}}

\def\omm0{\Omega_{\rm m,0}}
\def\omr0{\Omega_{\rm r,0}}
\def\oml0{\Omega_{\Lambda,0}}
\def\omk0{\Omega_{k,0}}

\numberwithin{equation}{section}

\def\be{\begin{equation}}
\def\ee{\end{equation}}
\def\bea{\begin{eqnarray}}
\def\eea{\end{eqnarray}}

\def\Mp{M_{\rm Pl}}
\def\calP{{\mathcal P}}
\def\calR{{\mathcal R}}
\def\calRk{{\mathcal R}_k}
\def\Mpbh{M_{\rm PBH}}
\def\Meq{M_{\rm eq}}
\def\fpbh{f_{\rm PBH}}
\def\rhotot{\rho_{\rm tot}}
\def\rhorad{\rho_{\rm rad}}
\def\rhopbh{\rho_{\rm PBH}}
\def\aeq{a_{\rm eq}}
\def\aform{a_{\rm form}}
\def\Mpc{\rm Mpc}
\def\erfc{\rm erfc}

\begin{document}


\title{Lecture notes on inflation and primordial black holes}

\author{Christian T.~Byrnes\footnote{E-mail: c.byrnes@sussex.ac.uk}}
\affiliation{Department of Physics and Astronomy, Pevensey II Building, University of Sussex, BN1 9RH, UK}

\author{Philippa S.~Cole\footnote{E-mail: p.s.cole@uva.nl}}

\affiliation{GRAPPA Institute\\\mbox{University of Amsterdam, 1098 XH Amsterdam, The Netherlands}\\}

\begin{abstract}

These lecture notes are based on those presented at the Theoretical Aspects of Astroparticle Physics, Cosmology and Gravitation School at the Galileo Galilee Institute in Florence in 2021\footnote{\url{https://agenda.infn.it/event/24368/}}. They aim to provide a pedagogical introduction and basic working knowledge of single-field inflation including the ultra-slow-roll regime, where the perturbations grow exponentially. This rapid growth is connected to the formation of primordial black holes (PBHs), a special dark matter candidate and probe of the initial conditions of the early universe. Although there are many textbooks and introductory texts about inflation, to the best of our knowledge there is no comparable introduction to ultra-slow-roll inflation. Furthermore, given their recent surge in popularity, there are numerous research articles and reviews on primordial black holes, however these notes aim to be more accessible for graduate students and those brand new to the topic. Some problems and solutions to primordial black hole-related calculations are also included. The reader of these lecture notes should come away being able to calculate the present-day abundance of primordial black holes produced from the density fluctuations left over at the end of single-field inflation with an ultra-slow-roll phase, and understand how this abundance compares with current observational constraints.


\end{abstract}

\maketitle


\newpage

\tableofcontents

\section{Introduction and the aims of these lecture notes}

In this short lecture course, we will aim to address two topics about two of the most fundamental questions in cosmology, what is the universe made of, and what were the initial conditions? We will study the leading theory of the early universe called inflation, and primordial black holes, which are a dark matter candidate.

We will first briefly study the historical motivations for inflation, before understanding how it can generate the initial density and temperature perturbations. Remarkably, inflation does this by invoking quantum mechanics - the theory of the smallest scales - and general relativity, to explain how the largest scale objects in the universe today are a consequence of physics on the smallest scales during the epoch of inflation. We will sketch the calculation of how the initial perturbations were generated and how to relate them to the perturbations observed at much later times via the cosmic microwave background and large-scale structure. 

There are many pieces of evidence that the majority of gravitationally clustered matter in the universe is non-baryonic (e.g.~galaxy rotation curves, structure formation and the bullet cluster), the so-called dark matter. 
Primordial black holes (PBHs) are the unique dark matter candidate which does not invoke a new particle to explain this. They can form within the first second after the Big Bang from the collapse of small-scale density perturbations. However, these perturbations need to be much larger in amplitude than those which we have observed on large scales with, for example, the Planck satellite. Therefore, non-standard initial conditions need to be invoked in order to produce a dramatic boost in the amplitude of density perturbations on smaller scales. Such non-standard initial conditions usually come in the form of a modification to the dynamics of the inflationary model. A study of PBHs therefore naturally relates to theories of inflation and provides a window onto the early universe on smaller scales than can be probed by any other means. Even if PBHs are only a negligible fraction of the dark matter, the detection of even a single PBH would have major implications for our understanding of the early universe, and current constraints on their existence are used to constrain the physics of inflation.

The suggestion of PBHs pre-dates inflation, but interest in them has never been higher since LIGO and Virgo's detections of gravitational waves caused by black hole mergers. Could some of those black holes be primordial?\footnote{{\tiny \url{https://www.quantamagazine.org/black-holes-from-the-big-bang-could-be-the-dark-matter-20200923/}}} 

Either of these subjects could easily be the subject of a much longer lecture course. We will therefore give only a summary of many of the details and derivations, with the goal being to provide a practical and modern knowledge of both subjects, with references provided to help you fill in the gaps. Because these are lecture notes, we are not aiming to provide a comprehensive list of references and will instead focus on review articles and books when possible. The emphasis will be on providing intuition and understanding rather than technical details, while the workshops led by Philippa Cole will be used to fill in some of the technical details and help build an understanding of some of the calculations involved. 

In these lectures we will use natural units where $c=\hbar=1$ unless otherwise stated, and use the reduced Planck mass which is related to Newton's gravitational constant by $\Mp^2=1/(8\pi G)$. 

\section{A crash course in background cosmology}

Although we will focus only on inflation and primordial black holes in these lectures, some knowledge of cosmology at the level of the homogeneous and isotropic background is essential before we can begin. If you have not studied much cosmology we highly recommend the very readable text book ``Introduction to Cosmology'' by Barbara Ryden which does not require any prior knowledge of general relativity.\footnote{\tiny{https://www.cambridge.org/core/books/introduction-to-cosmology/7E9E7C9C717570F1FFB3BA70F864A8FA}} 

Isotropic means that something looks the same in every direction, and this is true for the observed universe, as is best evidenced from observations of the cosmic microwave background. Homogeneous means the same everywhere, and 3D galaxy surveys confirm that statistically, on very large scales (scales larger than a cluster of galaxies), the universe is homogeneous. 

Although the universe today is very inhomogeneous on galactic scales, due to the fact that gravitational attraction has caused large structures to grow over billions of years, the early universe was much closer to being homogeneous on the scales which are large enough to be observationally probed today. Once again, the best probe comes from observations of the temperature differences in the CMB. Around 400,000 years after the Big Bang, the background temperature reduced sufficiently to allow electrons to bind to atomic nuclei in the epoch known as recombination\footnote{but which should really be called `combination' since electrons and atomic nuclei had never been bound before this time}. This meant that photons, which had previously been scattering with electrons and stopping them from binding to nuclei, were able travel long distances. Their final scatter with the electrons is defined by the `last scattering surface' and these newly free-streaming photons are known as the cosmic microwave background. The temperature fluctuations of the CMB that we observe today conserve an imprint of how overdense or underdense regions of the Universe were at the time of recombination, and therefore offer a window into an otherwise totally unobservable epoch. The density/temperature perturbations observed in the CMB have a characteristic amplitude of a few parts per hundred thousand, showing that the universe really was very close to being ``smooth'' at those times, and because gravitational attraction acts to make the density perturbation amplitude grow with time, it must be true that at even earlier times (at least on the scales which can be observed today) the universe was even closer to being perfectly smooth.

Three key equations describe the evolution of a homogeneous and isotropic universe, modelled in terms of the growth of the cosmic scale factor $a(t)$ which relates physical and comoving scales via
$$ r(t)=a(t) x, $$
where the comoving distance $x$ between two comoving (with the expanding universe) observers is constant, while the physical distance $r(t)$ grows proportionally with the expansion of the universe. The three equations are the {\bf Friedmann equation}
\be H^2\equiv \left(\frac{\dot{a}}{a}\right)^2=\frac{1}{3\Mp^2}\rho-\frac{\kappa }{R_0^2a^2}, 
\label{friedmann} \ee
the {\bf fluid equation}
\be \dot{\rho}+3 H\left(\rho+P\right)=0, \label{fluid} \ee
and the {\bf acceleration equation}
\be \frac{\ddot{a}}{a}=-\frac{1}{6\Mp^2}\left(\rho+3P\right). 
\label{acceleration} \ee
$\rho(t)$ is the energy density, $P(t)$ is the pressure, $\kappa$ is the curvature and the radius of curvature measured today is $R_0$. As usual a dot denotes a derivative with respect to cosmic time.

These are three of the most important equations in cosmology, so they deserve some explanation. We will not attempt to derive them, and a rigorous derivation requires a reasonable background in GR. Perhaps the first thing to stress is that only two of the three equations are independent, so it is possible (and a good exercise) to combine the first two equations to derive the third one. 

This means that we need a third equation to solve for the three unknowns, $a(t)$ (or equivalently $H(t)$), $\rho(t)$ and $P(t)$, even if we assume that we know the value of the curvature $\kappa$. The extra ingredient is the equation of state of the fluid(s) in the universe, i.e.~a relationship of the form
\be P=\omega \rho. \label{eos}\ee
In practise, we'll need multiple such relationships for each of the multiple components contributing to the energy density of the universe. Note that this linear relationship is obviously not the most general form of $P(\rho)$ possible, especially since we will assume that $\omega$ is a constant for each component. However, in practise, this simplifying assumption is perfectly adequate for describing our real universe during most of its evolution. It allows us to make substantial analytical progress in solving the Friedmann equation and hence enables us to determine how the different components of the universe evolved with time, as well as the history of the Hubble parameter.


We do not have time to derive the different properties of the different types of energy components of the universe, but to briefly summarise, the important components after inflation has ended are matter (which is pressureless, $\omega_{\rm mat}=0$), radiation ($\omega_{\rm rad}=1/3$), curvature ($\omega_{\kappa}=-1/3$) and the cosmological constant ($\omega_{\Lambda}=-1$).

By solving the fluid equation, component by component, you can (quite straightforwardly) derive
$$ \rho_{\rm rad}\propto a^{-4},\; \rho_{\rm mat}\propto a^{-3},\; \; \rho_{\kappa}\propto a^{-2}, \;\; \rho_{\Lambda}\propto a^{0}={\rm constant}. $$
This demonstrates that in the future, $\Lambda$ will dominate because it cannot be diluted, while in the past, radiation must have dominated. In between, we will see that there was a long period of matter domination. It is not believed that the curvature was ever important, except possibly before an epoch of inflation during the extremely early universe. But that is a different story which we will come back to near the end of this course. For now, it is important to note that treating the universe as if it only had one energy component is a useful exercise which provides a good approximation to the expansion of the universe during significant time periods of its history. We will go on to study such cases, before considering the more complete and complex case of a multi-fluid universe.

\section{Inflation}

Cosmologists can probe the epoch of the cosmic microwave background (CMB) around 400,000 years after the Big Bang `directly' by measurements of the photons which last scattered when the CMB formed. Less directly, we can extrapolate these measurements back in time until around 1 second after the Big Bang, to the epoch of Big Bang nucleosynthesis (BBN), and once again there is observational evidence that the predictions made for the abundances of the light elements are correct, giving us confidence that we understand the history of the universe back to a time when the energy scale was above an MeV and when neutrinos started free streaming. 

Going even further back in time is speculative, but there is good (albeit inconclusive) evidence that much less than a second after the Big Bang there was a period when the universe underwent accelerated expansion, popularly known as inflation. The three classic (and original) arguments for a period of inflation were that it could answer the following three riddles:

\begin{enumerate}

\item Why does the universe obey the cosmological principle, i.e.~why is it homogeneous and isotropic? Although this minimal assumption has turned out to be true, to the good fortune of cosmologists around the world, it is not easy to explain why the CMB temperature on opposite sides of the sky, which should never have been in causal contact, have nearly the same temperature. 

\item Why is the universe so close to being spatially flat? Since the effective energy density of curvature dilutes like $a^{-2}$ with the expansion, which is slower than both matter and radiation, it seems surprising that it didn't come to dominate the evolution of the universe before dark energy, e.g.~$\Lambda$, 
became important. 

\item Why don't we see any magnetic monopoles or other massive relics from the high energy early universe? If we keep extrapolating the laws of physics back to ever higher energy scales, then it appears ``likely'' that stable particles with a large mass would have formed in huge quantities. But such particles have never been seen, so how did they evade detection?

\end{enumerate}

Whilst the universe might be exactly flat and magnetic monopoles might simply never have existed, which resolves the latter two problems, it is extremely hard to resolve the horizon problem without an epoch of inflation. A description of how inflation resolves the three issues above is provided in any standard textbook on cosmology. For example, see the excellent description in ``Introduction to Cosmology'' by Barbara Ryden.

Today, most people research inflation because it is believed to have also generated the primordial perturbations, which are observed as temperature perturbations on the CMB sky. This is a remarkable story which potentially explains the origin of all cosmological structures - including galaxy clusters which are the biggest objects in the universe - as being due to quantum mechanical perturbations present during the brief period of inflation. There could not be a grander story of mighty oaks growing from tiny acorns than this one! 

\subsection{What is inflation?}

The simplest answer is that inflation refers to an early period when the expansion of the universe accelerated, meaning that $\ddot{a}>0$. Due to the cosmological constant $\Lambda$ (or more generally, dark energy), the universe also appears to be inflating now, but at a vastly lower energy scale (recall how much the temperature and the energy of the universe have decreased since BBN, and inflation must have occurred before BBN since otherwise the abundance of the primordial elements such as hydrogen and helium would have been diluted to nearly zero). 

From the acceleration equation, \eqref{acceleration}, we can see that $\ddot{a}>0$ means that the total equation-of-state parameter must satisfy 
$$ \omega<-\frac13 . $$ 
However, it turns out to be extremely difficult to get an extended duration of inflation\footnote{This does not mean that it lasts a long time, but rather that the universe grows by many orders of magnitude. We will soon learn a good way to quantify the amount of expansion during inflation.} unless the equation-of-state parameter satisfies $\omega\simeq-1$, so we will use this as a practical definition of inflation. The good news is that the Friedmann equations are quite straightforward to solve in this scenario, because the universe approximately behaves as if it was dominated by a cosmological constant.
However, inflation cannot be caused by the cosmological constant because the energy scale must be greater than the energy scale of the universe at the time of BBN, which is a vastly larger energy scale than the scale of dark energy/$\Lambda$ today. 
%
%
During inflation the scale factor grows at an exponential rate,
\be a(t)\propto e^{\Hinf t}. \label{eq:a-inf}\ee 

When discussing how much inflation is needed to fix the horizon, flatness and monopole problems, the question is not how long (in nano seconds, or any other units) inflation lasted, but rather by how much the scale factor grew during inflation, while $\omega\simeq-1$. A common and convenient measure of this growth is the efolding number, where 1-efolding means that the universe has grown by a factor of $e\simeq2.72$, 2-efoldings by a factor of $e^2$, and in general N-efoldings by a factor of $e^N$. The total number of efoldings of inflation is defined by
\be \Ninf \equiv \ln\left( \frac{a_{\rm inf,end}}{a_{\rm inf,initial}}\right) = \Hinf (t_{\rm inf,end}-t_{\rm inf,initial}) , \label{eq:Ninf} \ee
where the equality follows from \eqref{eq:a-inf}. By defining the duration of inflation in terms of the growth of the scale factor, we remove the degeneracy between the (unknown) energy scale of inflation (which is related to $\Hinf$) and for how long (in time) inflation took place.

\subsubsection{Length scales and the comoving Hubble scale}\label{sec:aH}

\begin{figure}
\centering{
\includegraphics[width=10 cm,clip=]{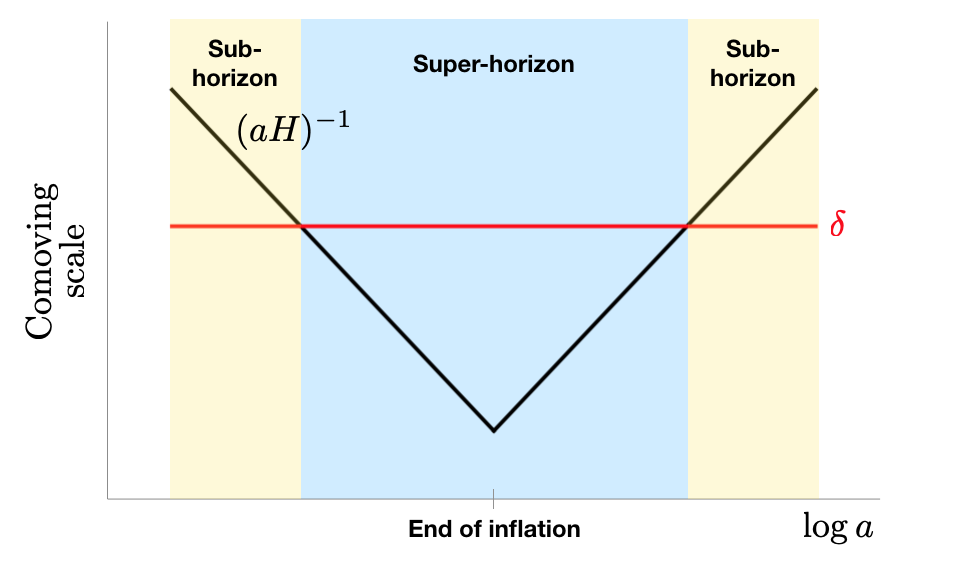}}
\caption{The evolution of the comoving Hubble horizon as a function of $N=\ln(a)$, showing how modes (with a constant comoving length) become super-Hubble ($k<aH$) during inflation and then re-enter the horizon after inflation has ended when $1/(aH)$ grows. Figure from \cite{Cole:2021cnu}.} \label{fig:Leach}
\end{figure}

We will soon go on to calculating properties of the perturbations generated during inflation. An important concept for this calculation will be the horizon scale, which is related to the Hubble scale and Hubble time. The full units for the Hubble constant measured today are the (somewhat weird-sounding) km/s/Mpc, meaning it has units of inverse time, and the Hubble time $1/H$ gives an estimate for the age of the universe. The Hubble distance, $c/H$ is the distance that light can travel in one Hubble time, making it a good estimate for the `scale of causal interactions' as a function of time. Recalling that physical scales are related to a comoving scale by a factor of $a$ and that we are using units with $c=1$, we can deduce that the comoving Hubble scale is given by
\be \frac{1}{aH}. \ee
During inflation $H\simeq\,$constant and hence $1/(aH)\propto 1/a$. During radiation domination $a\propto t^{1/2},\; H\propto1/t$ and therefore $1/(aH)\propto t^{1/2}\propto a$ while during matter domination $a\propto t^{2/3},\; H\propto 1/t$ and therefore $1/(aH)\propto t^{1/3}\propto a^{1/2}$. The evolution of the comoving Hubble scale as a function of efolding number $N=\ln(a)$ is shown in Fig.~\ref{fig:Leach}. Notice that even though the universe is today dominated by the cosmological constant, this has only become dominant in very `recent' times.

Because it is normal to measure the statistical properties of the primordial density perturbations in terms of the power spectrum in Fourier space, cosmologists often talk about `length' scales in terms of the comoving wavenumber $k\sim1/({\rm comoving\;length})$ which is usually measured in units of inverse megaparsecs (Mpc$^{-1}$). Notice that large values of $k$ correspond to small scales, and vice versa. An important concept for calculations of inflationary perturbations is whether the length scale is smaller or larger than the comoving Hubble scale. Perturbations with length scales which are larger than the comoving Hubble scale are referred to as super-Hubble or super horizon. This corresponds to $k<aH$, while a sub-Hubble mode satisfies $k> aH$. The mode crosses the Hubble scale when $k=aH$. As shown by Fig.~\ref{fig:Leach}, modes start off inside the Hubble scale, then exit during inflation and later re-enter the Hubble scale after inflation ends. Large-scale modes (corresponding to small values of $k$) exit earlier and re-enter later than small-scale modes. 
For reasonable choices of the inflationary energy scale, the number of efoldings between the moment our observable horizon scale exited the Hubble horizon during inflation and the end of inflation is between 50 and 60 \cite{Liddle:2003as}.

\subsection{What could cause inflation?}

We will start by considering a single scalar field $\phi$ with a potential energy given by $V(\phi)$. If the value of the potential of the scalar field is roughly constant and dominates the total energy density, then the background behaves like a cosmological constant, and hence drives the accelerated expansion. Because the background universe was close to being homogeneous and isotropic, we will make the assumption that the background inflaton field depends on time but is spatially invariant,
hence 
$\phi=\phi(t)$. The energy density and pressure of this field\footnote{This comes from the energy-momentum tensor of general relativity} are
\bea \rho_\phi=\frac12\dot{\phi}^2+V(\phi), \\
P_\phi=\frac12\dot{\phi}^2-V(\phi),
\eea
so we can see that the background equation-of-state parameter will be close to that of a cosmological constant, $w\simeq-1$, whenever the potential energy dominates over the kinetic energy of the field, i.e.~if
\be V(\phi)\gg \frac12\dot{\phi^2}. \ee 
Having an equation of state close to $-1$ means that the background density is only reducing very slowly, with the `extreme' case of $\omega=-1$ corresponding to a constant background energy. A good way to parametrise how quickly the background energy is diluting is via the slow-roll parameter
\be \epsilon_H\equiv -\frac{\dot{H}}{H^2}, \ee
and with a bit of algebraic manipulation using the (flat) Friedmann and acceleration equations we can write this in terms of the equation-of-state parameter as
\be \epsilon_H=-\frac{1}{H^2}\frac{\ddot{a}}{a}+1=\frac{3\Mp^2}{\rho} \frac{1}{6\Mp^2}\left(\rho+3P\right)+1=\frac12(1+3\omega)+1=\frac32(1+\omega). \ee
From the acceleration equation we can see that $\ddot{a}>0$ if $\omega<-1/3$, and therefore acceleration (which is the definition of inflation) corresponds to $\epsilon_H<1$. The commonly considered case of $\omega\simeq-1$ corresponds to $\epsilon_H\ll 1$.

\subsection{Slow-roll inflation}

We will now consider in more detail how to describe the (background) motion of a scalar field which causes inflation. The background equation of motion for such a field in an expanding background is given by\footnote{A derivation of this result goes beyond the scope of this course, but can be derived by applying the Euler-Lagrange equation to the action of the scalar field.} 
\be \ddot{\phi}+3H\dot{\phi}+V'=0, \label{eq:phi-eom} \ee
where the overdots refer to time derivatives and the prime denotes a derivative with respect to the inflaton field, $\phi$. The middle term represents the `Hubble friction' caused by the expansion of the universe, which would be zero if the background was static. 

Taking the time derivative of the Friedmann equation for a scalar field in a flat background
\bea H^2&=&\frac{1}{3\Mp^2}\left(V+\frac12\dot{\phi}^2\right), \\
\Rightarrow\, \, 3\Mp^2\times 2H \dot{H} &=&V'\dot{\phi}+\dot{\phi}\ddot{\phi} = \dot{\phi}\left(V'-(3H\dot{\phi}+V')\right)=-3H\dot{\phi}^2, \eea
where we used the equation of motion \eqref{eq:phi-eom}, we can hence rewrite the slow-roll parameter (without making any approximation) as
\be \epsilon_H\equiv -\frac{\dot{H}}{H^2} = \frac{\dot{\phi}^2}{2H^2\Mp^2}. \ee
This slow-roll parameter is therefore measuring the ratio of the kinetic and potential energies (up to order unity numerical factors) and hence inflation will occur when the kinetic energy is subdominant, as we had previously observed.

There are a huge number of ways of parametrising and classifying the slow-roll conditions. At lowest order in slow roll, two commonly used parameters are
\bea \epsilon_V &=&\frac{\Mp^2}{2} \left(\frac{V'}{V}\right)^2, \\
\eta_V&=& \Mp^2 \frac{V''}{V}, 
\eea
where again the primes denote differentiation with respect to the inflaton field $\phi$. The slow-roll conditions imply that
\be \epsilon_V\simeq\epsilon_H\ll1\, , \qquad |\eta_V|\ll 1, \ee
and one can show that these approximations are equivalent to dropping the following terms from the equations governing the evolution of the inflaton field
\bea \xcancel{\ddot{\phi}}+3H\dot{\phi}+V'=0 \\ H^2 =\frac{1}{3\Mp^2}\left(V+\xcancel{\frac12\dot{\phi}^2}\right) \eea
which results in the simplified slow-roll equations (which can be solved analytically for some simple choices of $V(\phi)$) 
\be 3H\dot{\phi}\simeq -V'\, \qquad 3H^2\Mp^2\simeq V. \ee



\subsection{Perturbation generation}

The reason for the continued interest in inflation four decades after it was first suggested has little to do with its ability to solve the classic horizon and flatness problems and is almost all to do with its ability to also generate the primordial perturbations. The full story of how this works is quite complex and wonderful. The basic idea is that the quantum mechanical uncertainty principle means that the inflaton field - and hence the universe - cannot become completely smoothed out by inflation. Quantum mechanical `seed' perturbations (which are normally only relevant on really tiny scales) become important on large, classical, scales because they are blown up due to the rapid expansion of the universe during inflation. We won't attempt to derive the quantum mechanical perturbations in these lectures, but simply state the results and give some explanations. For a detailed treatment, see e.g.~`The primordial density perturbation' textbook by Lyth and Liddle.

The relevant energy scale of inflation is given by the Hubble parameter\footnote{You may point out that the dimensions of the Hubble parameter are actually time$^{-1}$ which is true, but in natural units time (and length) scales are both the inverse of mass and energy scales.} and the typical amplitude of the scalar field perturbations ($\delta\phi$), as well as the metric (tensor) perturbations ($h$), are both linearly proportional to $H$. A derivation of this important result goes beyond the scope of these lectures. 

More precisely, we may write the power spectrum amplitude of the scalar (inflaton) and tensor (T) perturbations at horizon crossing during inflation as
\bea \calP_{\phi,*}=\left(\frac{H_*}{2\pi}\right)^2, \qquad \calP_{T,*}=\frac{8}{\Mp^2}\left(\frac{H_*}{2\pi}\right)^2\, . \eea

How do the power spectra of $\delta\phi$ relate to the observed CMB temperature perturbations? It turns out that this relationship is rather indirect. The quantity which is closely related to observations is rather the dimensionless curvature perturbation\footnote{I am skipping lots of details here, including the issue of gauges. The perturbed quantity $\delta\phi$ is a gauge dependent quantity but it turns out that $\calR$ is gauge independent, provided that it is defined carefully. There also exist numerous definitions and sign conventions for the curvature perturbation, which is often also denoted by $\zeta$, called `zeta'.} 
$$\calR =\frac{H}{\dot{\phi}}\delta\phi= \frac{1}{\Mp \sqrt{2\epsilon_H}}\delta\phi\simeq \frac{V'}{V}\delta\phi\simeq \frac{\delta V}{V}\simeq \frac{\delta\rho}{\rho}$$ 
and hence the power spectrum of $\calR$ measured at horizon crossing is given by 
\be \calP_{\calR,*}= \frac{1}{2\Mp^2\epsilon_{H,*}}\left(\frac{H_*}{2\pi}\right)^2, \ee
where $H$ and $\epsilon_H$ should be evaluated at horizon crossing when $k=aH$, but this time (or equivalently scale) dependence is often not shown explicitly.
%

\subsection{Observational tests of inflation}

The CMB temperature perturbations have been measured to high accuracy over three orders of magnitude in length scales, with the best constraints on most scales coming from the Planck satellite. Although the spectrum looks complicated, it is a remarkable fact that the statistical properties of the CMB map, which consists of about 10 million pixels, can be parametrised in terms of a primordial power spectrum which has only two free parameters, 
\be \calP_\calR=A_s \left(\frac{k}{\kp}\right)^{n_s-1}, \ee
where $A_s\simeq2\times10^{-9}$ is the amplitude of the primordial power spectrum and $n_s-1$ is the spectral index, with $n_s=1$ corresponding to a scale-invariant spectrum (i.e. $\calP_\calR$ independent of $k$). Observations favour $n_s\simeq1$ but exact scale-invariance has been ruled out with high significance. The pivot scale $\kp$ is not an additional free parameter but a normalisation scale, which is normally fixed to be in the `middle' of the range of data that can be constrained.

Given the definitions of $H$, the slow-roll parameters, and using $k=aH$, you can check (and this is a good exercise to do so - beware it is easy to get the second relation wrong by an overall minus sign) that
\bea \frac{d\ln H^2}{d\ln k}=-2\epsilon_H, \\ \frac{d\ln\epsilon_V}{d\ln k}=2\epsilon_V\left(2\epsilon_V-\eta_V\right), \eea
and hence to leading order in slow roll (which means we can approximate $\epsilon_H=\epsilon_V$) we find the following, very useful result for the spectral index
\be n_s-1\equiv \frac{d\ln \calP_\calR}{d\ln k} =-6\epsilon_V+2\eta_V. \label{eq:ns} \ee
To compare with CMB observations, the slow-roll parameters should be evaluated around the time when $\kp=aH$ during inflation, which corresponds to 50-60 efoldings before the end of inflation, as discussed in section \ref{sec:aH}. Although the slow-roll parameters vary slowly during slow-roll inflation and hence are essentially constant while the length scales observable in the CMB cross the (comoving) Hubble scale (which happens over the course of about 6-efolds of inflation), the slow-roll parameters can vary significantly between this time and times closer to the end of inflation. 

Equation \eqref{eq:ns} is an incredibly useful formula. It means that for any single-field slow-roll model of inflation, you can determine the scale dependence of the primordial power spectrum in terms of the derivatives of the potential with respect to the scalar field and nothing more than that. This normally makes the calculation quite straightforward. However, one needs to know at which scale (or equivalently, value of the inflaton field) to evaluate the slow-roll parameters.

We can find the relation by starting with the formula $N\propto H t$ during inflation, and we will use $t_*$ to denote the time when $k$ equals the comoving Hubble scale, $k=aH$, while $t_e$ denotes the end of inflation. We therefore have
\be N=\int_{t_*}^{t_e}H dt=\int_{\phi_*}^{\phi_e}\frac{H}{\dot{\phi}}d\phi \simeq \frac{1}{\Mp^2}\int_{\phi_e}^{\phi_*} \frac{V}{V'}d\phi, \ee
where one needs to use the slow-roll equations of motion in order to get to the final result. It is often a good approximation to use $\phi_*\gg\phi_e$ for values of $N\gg1$.

The amplitude of the primordial tensor perturbations is often quoted as a ratio compared to the amplitude of the scalar perturbations, 
with the tensor-to-scalar ratio defined as
\be r\equiv \frac{\calP_T}{\calP_\calR} =16\epsilon_H . \ee 
Although this quantity also depends on scale (and should again be evaluated at horizon crossing), $\epsilon_H$ varies slowly and $r$ is constrained to be small (but it has not yet been detected), so the mild scale dependence of $r$ is normally unimportant. \\

{\bf The Planck constraints on inflation} have determined that 
$$n_s-1=0.965\pm0.004, \qquad r\lesssim 0.1 $$
while the addition of ground-based data from Bicep-Keck strengthens the constraint on the tensor-to-scalar ratio to be $r<0.044$ at the 95$\%$ confidence level \cite{Akrami:2018odb}. This implies that $\epsilon_H\simeq\epsilon_V<r/16\lesssim 0.003$ and therefore we can see from the formula for the spectral index, \eqref{eq:ns}, that the scalar power spectrum deviates too far away from being scale-invariant for the deviation to be due to the $\epsilon$ slow-roll parameter, and hence the data requires $\eta_V<0$, which implies that $V''<0$. The observational constraints on $\epsilon_V$ and $\eta_V$ are shown in figure 6 of the arXiv version of \cite{Akrami:2018odb}, and this demonstrates that even with such a limited number of non-zero inflationary parameters we can deduce something very non-trivial about the inflaton potential. However, all of this analysis has been conducted assuming the simplest case of single-field slow-roll inflation.




\newcounter{contribution}

\section{Ultra slow-roll inflation}

Recall the equation of motion for the inflaton field
\be \ddot{\phi}+3H\dot{\phi}+V'=0, \ee
where the prime in this equation refers to a derivative with respect to $\phi$ and the dot refers to a derivative with respect to (cosmic) time. The normal SR approximation is to drop the $\ddot{\phi}$ term and hence turn the second order equation of motion into a (much simpler) first-order equation, giving the result $3H\dot{\phi}\simeq-V'$. This approximation is normally valid provided that $\epsilon_V\ll1 \, \Rightarrow\, V'\ll\Mp V$, but not always. 

What happens if $V'=0$? Then the SR approximation would say that $\dot{\phi}=0$ implying that the field is not rolling at all. This does not imply that inflation has ended, rather the opposite. If the inflaton field is not at the bottom of the potential then the field will not move again and hence the universe is dominated by the potential energy of the inflaton field, which at least classically is the same as a cosmological constant and hence this gives rise to eternal de Sitter expansion (called eternal inflation). We clearly do not live in such a universe. 

However, if $V'=0$ then it would be wrong to neglect $\ddot{\phi}$ in comparison to $V'$, and we should instead study the following equation of motion:
\be \ddot{\phi}+3H\dot{\phi}=0. \label{eq:phi-USR} \ee
This is often described as the equation of motion for ultra-slow-roll (USR) inflation. Why USR? Because the field velocity decreases extremely quickly during USR inflation. We can solve \eqref{eq:phi-USR} by using the substitution $v=\dot{\phi}$ and recalling that $H=\dot{a}/a$ to find the solution
\be \dot{\phi}\propto a^{-3} \propto e^{-3N} \ee
which shows that the kinetic energy (KE) decreases like
$$ {\rm KE} \propto \dot{\phi}^2\propto a^{-6}. $$
Hence the potential energy very quickly dominates over the KE, but $\dot{\phi}$ remains important both in order to get the inflaton past the flat part of the potential so as to avoid eternal inflation and also in order to produce the inflaton perturbations. 

Note that the first slow-roll parameter
\be \epsilon_H = \frac{\dot{\phi}^2}{2\Mp^2 H^2}\propto a^{-6} \ll1 \ee
is small and rapidly decreasing during inflation, which implies that $H$ is very close to constant. In contrast, $\epsilon_V=0$ precisely if $V'=0$, so whilst both versions of the epsilon slow-roll parameter are small during USR inflation, they do not have the same order of magnitude or time dependence. Hence, unlike in SR inflation, one cannot use them interchangeably. 

We now introduce a ``second'' slow-roll parameter which measures how quickly $\epsilon_H$ varies with time,
\be \eta_H\equiv \frac{\dot{\epsilon_H}}{H\epsilon_H} \simeq\frac{\frac{d\epsilon_H}{dN}}{\epsilon_H} \ee
where we have used $N\simeq Ht$ to get to the final equality, which is valid during inflation. Just as $\epsilon_H\sim\epsilon_V$ in the slow-roll approximation, $\eta_H\sim\eta_V$ in the slow-roll approximation too\footnote{We caution the reader that there exist a huge number of different slow-roll parameters in the literature, some of which vary by definition and some by notation. The subscripts are often not used.}. We can hence see the important result that $\eta_H\simeq0$ during SR inflation, while $\eta_H\simeq-6$ during USR inflation. 

\begin{figure}
\centering{
\includegraphics[width=10 cm,clip=]{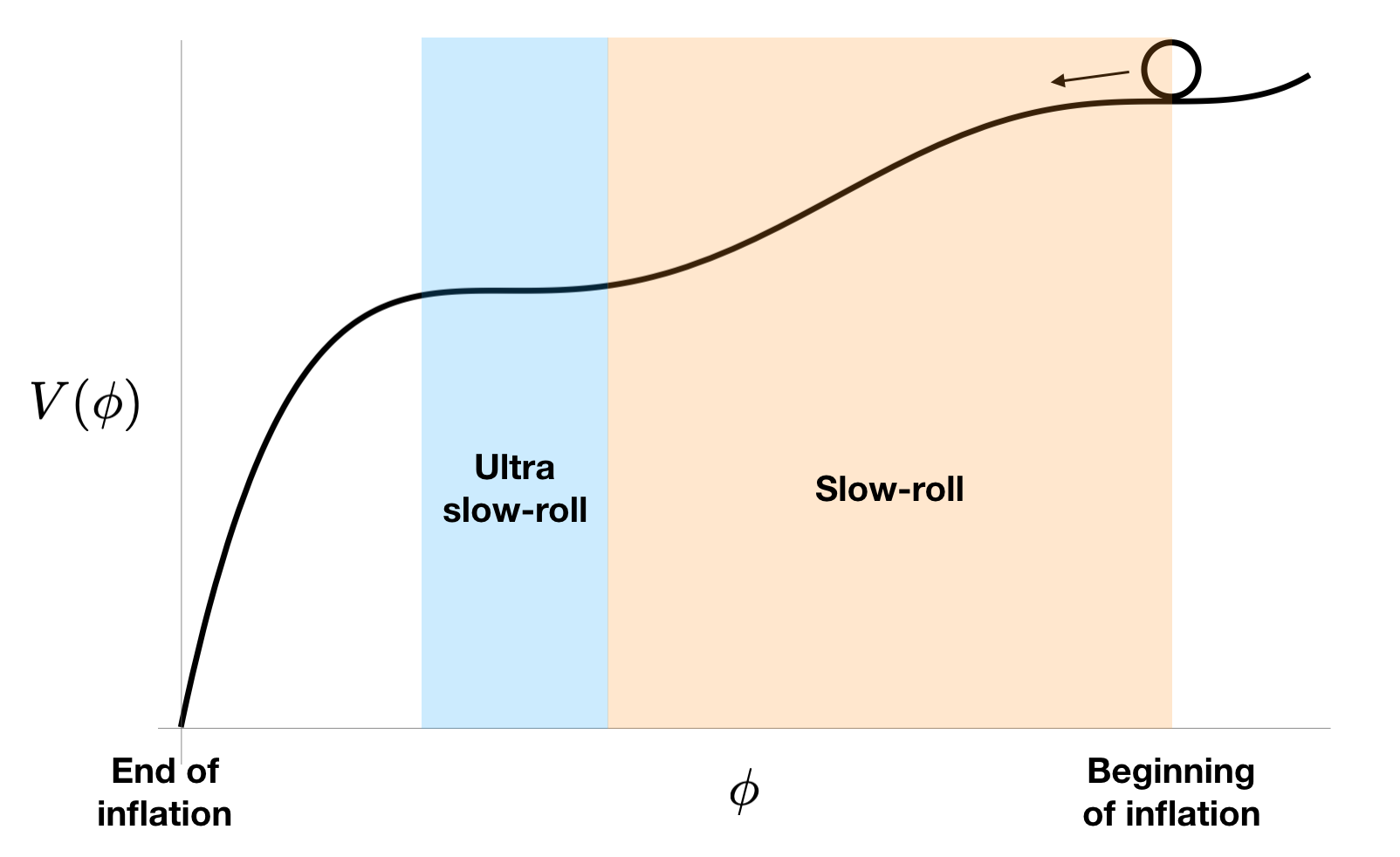}}
\caption{An example of an inflaton potential that includes an inflection point. Figure from \cite{Cole:2021cnu}.} \label{fig:inflection}
\end{figure}

In Fig.~\ref{fig:inflection} we show an example of a potential which has an inflection point, meaning that the derivative of the potential is zero at this field value. USR inflation will occur while the inflaton field traverses the inflection point and for a brief period either side. 

\subsection{The amplitude of the perturbation}

The standard formula for the amplitude of the perturbations generated during single-field slow-roll inflation is
\be \calP_\calR=\frac{H^2}{8\pi^2\Mp^2 \epsilon}. \label{eq:calP} \ee 
It turns out that this remains partially true even when USR inflation occurs (only partially because it is not valid on all scales), but subject to some very important caveats. As we will see later, during USR inflation the perturbations do not freeze out at horizon exit, and hence the formula cannot be evaluated at horizon crossing but rather only after USR inflation ends. Another key difference is that while SR inflation implies that $\epsilon_V\simeq\epsilon_H$ and hence you can choose to use either definition of $\epsilon$, during USR the two become extremely different and indeed $\epsilon_V=0$ if $V'=0$, but this does not imply that $\calP\rightarrow\infty$. In the formula for the power spectrum, \eqref{eq:calP}, must be evaluated using $\epsilon=\epsilon_H$. 

Recalling that $\epsilon\propto a^{-6}$ during USR inflation we can see that $\calP_\calR\propto e^{6N}$ since $H$ is very close to constant, i.e.~the perturbations (at least on some scales) grow very rapidly. This should make it clear that USR inflation cannot last very `long', i.e.~for a large number of efolds of inflation, because if it did then $\calP_\calR\rightarrow 1$ would be reached. At this point, not only would perturbation theory break down since the perturbations are no longer small, but more seriously, $\calP_\calR\rightarrow 1$ also corresponds to the onset of eternal inflation. This occurs when
$$ \sqrt{\calP_\calR}\sim \frac{H}{\sqrt{\epsilon}}\sim \frac{H}{\frac{\Delta\phi}{\Delta N}}\sim 1$$
which means that the quantum fluctuations of the field -- which have amplitude $\delta\phi\sim H$ -- have the same magnitude as the distance which the inflaton field will (at the background level) roll down the potential during 1 efold of inflation, $\Delta\phi\sim H\Delta N$. At this point the inflaton field is about as likely to quantum mechanically ``jump'' up the potential (meaning that it moves backwards towards the starting point) as it is to continue moving towards the minimum of the potential where inflation can end. Hence inflation will never end (at least not in all locations) and this corresponds to the regime of eternal inflation. \\

{\bf Key messages:}
\begin{enumerate}
\item You have to be very careful calculating the perturbations during USR inflation
\item An inflection point can lead to eternal inflation: It is then important to check that the inflaton field has enough kinetic energy to get past the flat part of the potential and avoid eternal inflation
\item If $V'\neq 0$ exactly then USR inflation normally won't last long, because $\dot{\phi}$ is decreasing so quickly and hence slow roll with $3H\dot{\phi}\simeq V'$ will typically be quickly reached. 
\end{enumerate} 

The (quasi) scale invariance of the perturbations generated during inflation is often intuitively explained as being a consequence of the (quasi) time translation invariance of the quasi de Sitter space expansion caused by inflation. Whilst these things do correspond to $H\simeq$\,constant and hence both $\delta\phi$ and the tensor perturbations are nearly scale invariant, the scale invariance of the curvature perturbation is not guaranteed. During SR inflation the near constancy of $\epsilon$ implies via \eqref{eq:calP} that the curvature perturbations are also nearly scale invariant, but when $\epsilon_H$ varies, scale invariance is normally lost. However, and quite remarkably, USR inflation can give rise to a scale-invariant spectrum of perturbations as well.

{\bf Warning:} Do not confuse the time dependence with the scale dependence of the primordial curvature perturbations during USR inflation. We can only observe (indirectly) the primordial power spectrum after inflation has ended, so we just see its scale dependence evaluated at this time. For purely SR inflation it is enough (and much simpler) to just evaluate the primordial power spectrum at the time when modes exit the horizon during inflation, because the perturbations thereafter have a constant amplitude so we can stop tracking their evolution. During USR inflation this is not true, for reasons which will be explained in the next subsection. 

\subsection{USR perturbations on large scales}

We will now study more quantitatively the evolution of the curvature perturbation, by starting with (but not deriving) its equation of motion. This is most conveniently written and solved in terms of conformal time $\tau$,\footnote{Conformal time is just as often called $\eta$ in the literature, but we are already using $\eta$ for one of the `SR' parameters.} which is related to cosmic time by the scale factor with
$$ a d\tau=dt. $$
The name ``conformal time'' comes from the fact that the scale factor $a(t)$ becomes a conformal (overall) factor of the Friedmann-Le Maitre-Robertson-Walker metric when written in terms of this time coordinate. This is true for any global geometry (flat, closed or open) but specialising to the flat case for simplicity we can see this explicitly from
\bea ds^2 &=& -c^2 dt^2+a^2(t) \sum_{i=1}^3 dx_i^2 \\ &=& a^2(\tau)\left( -c^2d\tau^2+ \sum_{i=1}^3 dx_i^2 \right). \eea
In terms of conformal time, the equation of motion of the curvature perturbation, $\calR(k)\equiv\calRk$, is
\be \frac{\partial^2 \calRk}{\partial \tau^2}+2\frac{\frac{\partial z}{\partial \tau}}{z} \frac{\partial \calRk}{\partial \tau}+k^2 \calRk=0, \label{eq:calRk} \ee
where
\be z^2=2a^2\Mp^2\epsilon_H. \ee
In the $k\rightarrow0$ limit one can find a partially analytic solution by first substituting $v_k=\partial \calRk/\partial \tau$, solving for $v_k$ and then integrating, with the general solution (which is written in terms of two $k$ dependent constants that depend on the initial conditions) given by
\bea \calR_{k\rightarrow0}&=& C_k+D_k \int^{\tau}\frac{d\tau'}{a^2\epsilon_H} \\ &=& C_k+D_k \int^t \frac{dt'}{a^3\epsilon_H}.\eea
The constant mode $C_k$ corresponds to the same mode as we see in slow-roll inflation which remains constant after horizon exit, while $D_k$ corresponds to the mode usually called the decaying mode. During SR inflation $\epsilon\simeq$ constant and hence the decaying mode does decay like $a^{-3}$, showing that in this case $\calRk$ does indeed freeze out at around the time of horizon crossing, after which $k\ll aH$.\footnote{Note that while this statement is true, this has not been shown rigorously - like many steps in these lecture notes - because we have only solved \eqref{eq:calRk} for $k=0$ rather than finding the general solution and then taking the $k\rightarrow0$ limit.} For constant values of $\eta$ it is straightforward to show that $\epsilon\propto a^\eta$ and therefore the ``decaying'' mode will in fact grow for $\eta\leq-3$. In particular, during USR inflation we have $\eta=-6$ and therefore
\bea \calRk\propto D_k \int^t dt' a^3 \propto \int^t dt' e^{3Ht'}\propto e^{3Ht}\propto a^3, \eea
where we have used the fact that $a\propto e^{Ht}$ during inflation. Hence we see that during USR inflation the perturbations have a time dependence which grows like 
$$ \calP_\calR^{\rm USR}\sim \calRk^2\propto a^6 $$
rather than freezing out at horizon exit, as we had previously argued based on \eqref{eq:calP} and the behaviour of $\epsilon_H\propto a^{-6}$ during USR inflation. 

Note that it is important to clearly distinguish the notions of the time-dependent power spectrum and the $k$-dependent power spectrum of the comoving curvature perturbation. Choosing a single scale, tracking the time-dependence of the comoving curvature perturbation illuminates the superhorizon growth of the perturbations. Meanwhile, choosing a single time (usually the end of inflation) to evaluate the power spectrum across all scales quantifies the corresponding final amplitude of the perturbations at the end of inflation.


\section{Primordial black holes}

It is important to realise that the ample evidence for dark matter does not just come from observations of the `late' universe, such as the observations of galaxy clusters, galactic rotation curves and the bullet cluster, but also from the growth of perturbations between the time when the CMB formed and the time when the first galaxies started to form, as well as consistent measurements of the baryon-to-photon ratio from the time of BBN about 1 minute after the Big Bang and the CMB formation about 400,000 years later. Hence, dark matter must have already existed during the early universe and this proves that `stellar' black holes which form from the collapse of stars are not a viable DM candidate. However, PBHs form very early (we later determine the relation between their mass and formation time) and hence PBHs could be the DM. We will summarise the observational constraints on the PBH abundance and conclude that there remains at least one mass window (approximately the mass of an asteroid) in which all the of the DM could be made up of PBHs. More massive PBHs could form a subdominant component of DM. PBHs are a special DM candidate both in terms of their large mass and because they uniquely do not require the existence of a new particle. However, they instead require special initial conditions to form.


\subsection{PBH formation}

Recall the discussion in section \ref{sec:aH} about the comoving Hubble scale $1/(aH)$. Modes with a constant comoving length scale $\sim1/k$ are initially smaller than the comoving Hubble scale, such that $k>aH$. However, these modes exit the horizon at the time that $k=aH$, and afterwards will then have a longer wavelength than the comoving Hubble scale which continues to shrink, meaning that $k<aH$ until after inflation ends and the comoving Hubble scales starts to grow. While $k<aH$ the mode is described as being `super-horizon' or `super-Hubble', and after inflation ends the modes will at some point `re-enter' the horizon when $k=aH$ again. The significance of horizon entry is that this time corresponds to when the corresponding horizon scale comes into causal contact, i.e.~the time when information travelling at the speed of light can travel across the comoving scale $1/k=1/(aH)$ in one Hubble time, $1/H$.

The horizon scale is a key concept in PBH formation because it is a causal process. Gravity (which travels at the speed of light) needs to communicate the existence of an overdensity in order for gravitational collapse to begin. Therefore, a PBH of scale $1/k$ cannot form while $k<aH$. PBHs form with a mass comparable to the horizon mass $M_H$, which means that there is an approximate 1-2-1 relation between the PBH mass $\Mpbh\sim M_H$, PBH scale $k$, and time of formation. We know that before matter domination began about 50,000 years after the Big Bang the universe was radiation dominated, and that it was radiation dominated at the time of BBN when the primordial elements (mainly hydrogen and helium) were formed about a minute after the Big Bang. At even earlier times we cannot be sure, but in the standard model of cosmology the universe was radiation dominated from shortly after the time when inflation ended and the inflaton field reheated the universe by decaying into radiation.\footnote{For a review article about the equation of state of the early universe see \cite{Allahverdi:2020bys}.} The horizon mass at the time of radiation-matter equality was an enormous $M_H=\Meq\sim 10^{16}M_\odot$ and we will therefore focus on PBH formation during radiation domination during these lectures, when the horizon mass was smaller.

The background pressure is very large during radiation domination, $P=\omega\rho=\rho/3$, meaning that only large amplitude perturbations will have a strong enough gravitational attraction to overcome the pressure forces and collapse into a black hole. The typical amplitude of the density perturbations on CMB scales is
$$ \delta=\frac{\delta\rho}{\rho}\sim \calR\sim \sqrt{t\calP_\calR}\sim\sqrt{A_s}\sim5\times10^{-5}, $$
which is far too small to lead to PBH formation. To form a PBH we instead need $\delta\rho/\rho\sim1$ at the time of horizon entry and hence PBH formation requires special initial conditions.  We know that the spectral index satisfies $n_s-1\simeq0$ on CMB scales, and if this extends to the smallest scales then zero PBHs will form of any mass.\footnote{Note that there are other mechanisms which could form PBHs, for example the collisions of cosmic strings. However, all of these alternatives require new (beyond the standard model) physics to form, and we won't consider them further. For a brief review see \cite{Green:2014faa}.}

The original estimate for the collapse threshold for PBH formation was made by Bernard Carr in 1975 (while he was Hawking's PhD student) using the Jean's length and time and using Newtonian gravity. He found that an overdensity would collapse if
\be \delta\equiv\frac{\delta\rho}{\rho}|_{k=aH}>\delta_c=c_s^2, \label{eq:Carr} \ee
where $c_s$ is the sound speed of perturbations, which is an important quantity because this determines how quickly a pressure wave caused by the overdensity can travel from the centre to the edge of the perturbation. During radiation domination, $c_s=1/\sqrt{3}$ so $\delta_c=c_s^2=\omega=1/3$. Both one-dimensional, and more recently three-dimensional GR simulations \cite{Yoo:2020lmg,Escriva:2021aeh}, have shown that 
\be \delta_c\simeq0.45, \ee
which is quite close to Carr's original estimate. They have also shown that the collapse threshold has only a mild dependence on the initial density profile. However, \eqref{eq:Carr} does not remain accurate in the limit of a matter dominated universe with $c_s\rightarrow0$ because his estimate assumed the initial overdensity was spherically symmetric. Whilst this is a good approximation for rare peaks in the density field \cite{Bardeen:1985tr}, this is not valid for small values of $c_s$ where overdensities can more easily collapse due to the lack of pressure, and it is the non-sphericity of smaller overdensities which prevents everything from undergoing gravitational collapse.

We would now like to estimate the amplitude of $\calP_\calR$ which can generate an `interesting' number of PBHs. As is normal for models of DM production, we only need a tiny fraction of the total energy density to be in the form of PBHs at the time of PBH formation in order to get a significant fraction of the DM to consist of PBHs today. This fraction is normally parametrised by 
\be \fpbh\equiv \frac{\rho_{\rm PBH}}{\rho_{\rm DM}}, \ee
and measured today, where $\fpbh=1$ means that all of the DM is made out of PBHs and therefore DM would not be a new particle.

During radiation domination $\rhotot=\rhorad\propto a^{-4}$ while after formation $\rhopbh\propto a^{-3}$ and therefore the PBH fraction $\rhopbh/\rhotot$ grows proportionally to the scale factor $a$ from the time of formation until matter-radiation equality. Denoting the fraction of the universe's energy density in PBHs at the time of formation as $\beta$, we therefore have
\be \fpbh=\frac{\rhopbh}{\rho_{\rm DM}}|_0\simeq \frac{\rhopbh}{\rhotot}|_{\rm eq}\simeq \frac{\aeq}{\aform}\beta. \ee
PBH formation is `fast' because the initial over density is so huge, unlike galaxy formation which begins with tiny initial overdensities of order $\delta_{\rm initial}\sim 10^{-4}$, and hence takes  billions of years. PBH formation takes about ten Hubble times, meaning that the collapse occurs at a time $10/H_{k=aH}$ where $1/H_{k=aH}$ is the Hubble time when the mode re-entered the horizon. Notice that this corresponds to the universe growing by only about e-folding during radiation domination, because $a\propto t^{1/2}$ during this time and therefore $\Delta N_{\rm PBH \;formation}\sim \ln(10)/2\sim1$. We can therefore approximate the time of PBH formation as being equal to the time when the mode of the overdensity which will form the PBH re-enters the horizon, i.e.~the time when $k=aH$.

As stated before, the PBH mass is comparable to the horizon mass at the time of formation, so 
\be \Mpbh\sim M_H=\rho V=\frac43\pi\rho \left(\frac{1}{H}\right)^3\propto \rho^{-1/2}\propto a^2\propto t \ee
where it is important to realise we used the physical (not comoving) Hubble scale $1/H$ as an estimate of the radius of the horizon volume $V$ and we also used $H^2\propto\rho\propto a^{-4}$ to derive some of the relations in the above equation. Note that $\Mpbh\propto a^2$ rather than $a^3$ because the density is decreasing while the horizon volume increases. Inserting the numerical factors one can find
\bea \Mpbh &=& \left(\frac{\aform}{\aeq}\right)^2 \Meq \simeq \left(\frac{\aform}{\aeq}\right)^2 10^{16} M_\odot, \\ 
\Mpbh&\sim& 10^{15}{\rm g}\, \frac{t}{10^{-23}{\rm s}}. \eea
The reason why the mass-to-formation time is often written in the form seen above\footnote{For example, see the very first equation in the PBH review article by Green and Kavanagh \cite{Green:2020jor}.} is because PBHs with an initial mass of $10^{15}$g will be evaporating today due to Hawking evaporation of the black hole. Those which form with a smaller initial mass will have already evaporated completely, and we can neglect the impact of Hawking radiation for BHs which form with a significantly larger initial mass because their evaporation timescales are greater than the age of the universe.

In terms of the comoving wavenumber $k$ measured at horizon entry after inflation, 
\be k=aH\propto t^{1/2}\propto a^{-1} \;\; \Rightarrow \;\; M_H\propto k^{-2} \ee
and inserting numerical factors leads to 
\be \Mpbh\simeq M_H\sim 10^{13}M_\odot \, k^{-2} \, \Mpc^2. \ee 

For the case of a solar mass PBH, $\Mpbh=M_\odot=2\times10^{33}$g, we can make order of magnitude estimates that they form when
\be k\sim 10^7 \Mpc^{-1}, \;\; t\sim 10^{-6} {\rm s}, \;\; \aform\sim 10^{-8}\aeq. \ee
The corresponding energy at this time is about $200$ MeV which corresponds to the time of the QCD transition when quarks bind into hadrons. We will see the significance of this coincidence later. Given that $\aform\sim 10^{-8}\aeq$, only 10 parts per billion of the universe needs to be in the form of PBHs at the formation time (i.e.~$\beta\sim10^{-8}$) in order for all of the DM to be made out of solar mass PBHs.

\subsection{Estimating the collapse fraction $\beta$}\label{sec:beta}

\begin{figure}
\centering{
\includegraphics[width=14 cm,clip=]{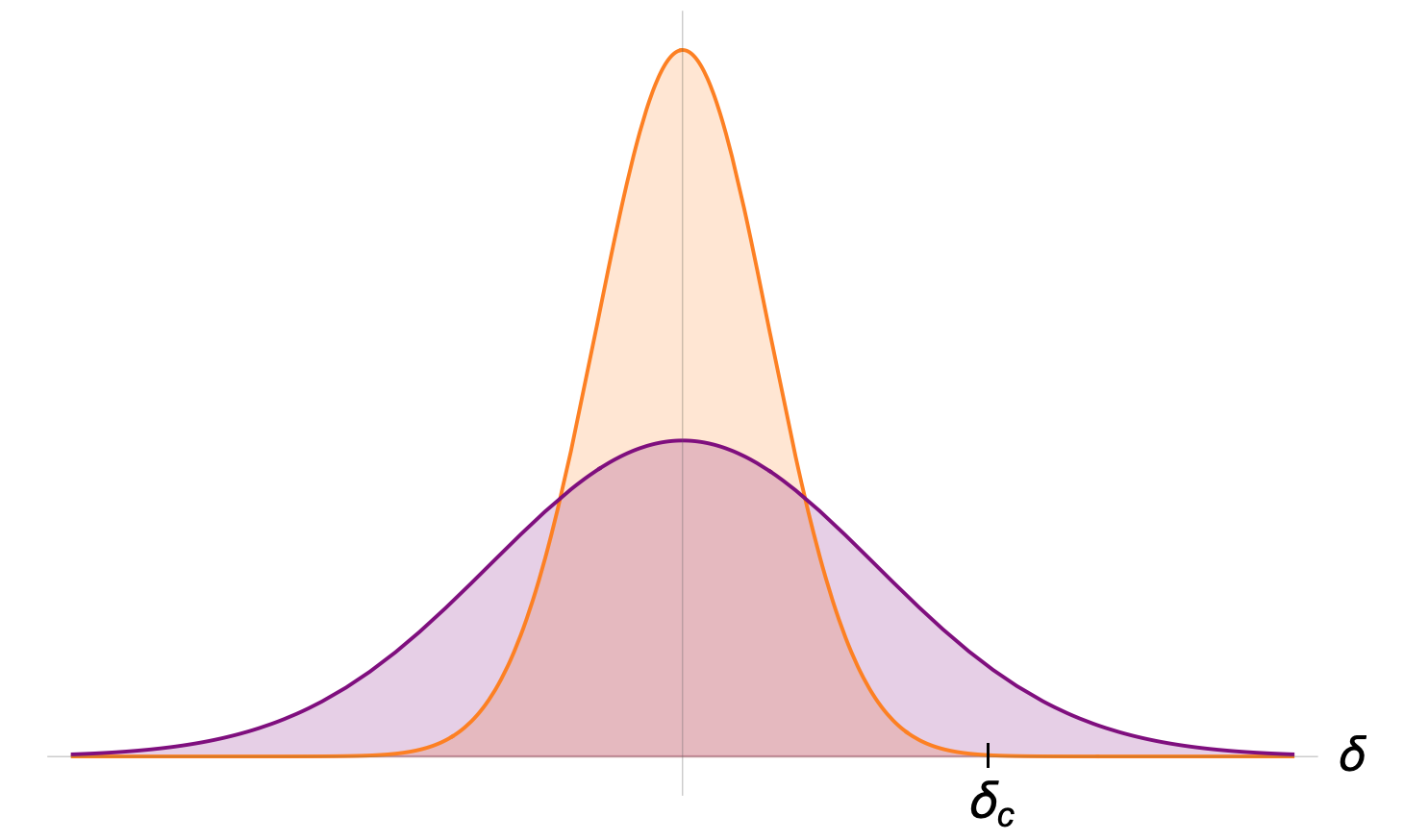}}
\caption{Two PDFs with different variances describing an initial Gaussian density field at a particular scale. The purple curve has a larger variance, and therefore the integral under the curve above the critical threshold density is larger, resulting in a larger abundance of PBHs. The variance is related to the amplitude of the primordial power spectrum at the same scale.} \label{fig:gauss}
\end{figure}

The easiest method is to use the Press-Schechter formalism, in which the collapse fraction of the universe into PBHs at the time of formation (or horizon entry of the relevant mode) is estimated by calculating the fraction of the universe with $\delta>\delta_c$,
\be \beta(\Mpbh)=\frac{\rho(\Mpbh)}{\rhotot}|_{\rm formation}=\int_{\delta_c}^\infty P(\delta) d\delta, \ee 
where $P$ is the pdf, not to be confused with the power spectrum. For example, if we assume an initial Gaussian density field, then for each scale, we can draw the PDF of the fluctuations as in figure \ref{fig:gauss} where two PDFs with different variances are shown. The PDF describes how likely it is that a given fluctuation will have density $\delta$. The probability of an overdensity being larger than the critical threshold for collapse is greater for a density field with a larger variance (purple curve), and hence the abundance of PBHs will be larger because the integral of the PDF between $\delta_c$ and infinity will also be larger. The variance of the field at each scale is related to the amplitude of the primordial power spectrum as we will now see, meaning the greater the amplitude of the power spectrum, the more PBHs produced.

For simplicity in these lectures, we will assume $\calR=\delta$ at horizon crossing and hence both quantities have the same power spectrum, although we caution that this is not a very accurate approximation. In reality, $\delta$ has a different scaling with $k$, and they are related at linear order by 
$$ \delta = \frac49 \left(\frac{k}{aH}\right)^2 \calR, $$
during radiation domination,\footnote{sometimes there is also a minus sign difference, due to differing definitions of $\calR$ used in the literature, but this minus sign has no impact on the power spectrum.} which shows that at horizon entry (when $k=aH$) they are related by an order unity factor. 

Technically one should also use a window function to smooth the density contrast $\delta$ on the scale of PBH formation, $R\sim1/k$, whose variance is related to the primordial power spectrum of the density contrast by
\be \sigma^2(R)=\int_0^\infty \widetilde{W}^2(kR)\calP_\delta(k) d\ln k, \ee
where $\widetilde{W}$ is the Fourier transform of a real space window function and $\calP_\delta(k)$ is the dimensionless power spectrum of the dimensionless density perturbation.

For simplicity we will neglect these complications and instead use 
$$ \sigma^2=\calP_\calR $$
as a rough estimate. Then if we assume that the perturbations are Gaussian distributed with variance $\sigma^2$, and using the fact that $\beta\ll1$, we can make an asymptotic expansion in the limit $\delta_c/\sigma\gg1$ to show that
\be \beta \simeq \frac12 \erfc\left(\frac{\delta_c}{\sqrt{2}\sigma}\right)\simeq \frac{\sigma}{\sqrt{2\pi}\delta_c} e^{-\delta_c^2/(2\sigma^2)}. \ee 
As a {\bf very} rough estimate, we can invert this to estimate
\be \calP_\calR\sim\sigma^2\sim\frac{\delta_c^2}{\ln(1/\beta)}\sim\frac{0.2}{\ln(1/\beta)}. \ee
Notice that $\calP_\calR$ is only logarithmically sensitive to $\beta$, whilst instead $\beta$ is exponentially sensitive to the power spectrum amplitude. Hence, relatively small changes to $\delta_c$, or the relation between $\calR$ and $\delta$, or changes to the choice of the window function all lead to exponentially large changes in $\beta$. If we manage to measure the value of $\fpbh$, and hence determine $\beta$, then there is freedom to slightly adjust the amplitude of the power spectrum (which relates to the duration of the non-slow-roll phase during inflation) to match the required abundance. However, if we measure the amplitude of the power spectrum very accurately with non-PBH probes, then the value of $\beta$ will consequently be very finely constrained despite the fact that it could have varied by orders of magnitude with even a slightly different measurement of the power spectrum amplitude.

Consider again the case of solar mass PBHs and assuming that $\fpbh=1$, it is straightforward to estimate that the required power spectrum amplitude is
\be \calP_\calR\sim\frac{0.2}{\ln(10^8)}\simeq0.011\sim10^{-2}. \ee
For much smaller PBH masses the relevant values of $\beta$ also become much smaller (due to the longer period of expansion between PBH formation and radiation-matter equality). The tightest observational constraint on $\beta$ is $\beta\lesssim 10^{-28}$ for $\Mpbh\sim10^{15}$g \cite{Carr:2009jm}, i.e.~the PBH initial mass which corresponds to them decaying today and hence having a large observational signature through Hawking radiation. How much impact does this twenty orders-of-magnitude tighter constraint have on the consequent constraint on the power spectrum amplitude? The answer is not much, only by a factor of $28/8=3.5$. Hence the range of power spectrum amplitudes which are of interest for PBH formation is quite limited, lying in the range $\calP_\calR\sim10^{-3}-10^{-2}$, independently of the PBH mass and for any potentially observable value of $\fpbh$. However, primordial non-Gaussianity or an early matter dominated epoch can strongly change the constraints on the power spectrum amplitude, but a study of these topics goes beyond the scope of these lectures.

How sensitive are these results to changes in the collapse threshold $\delta_c$? Phrased in terms of $\beta$ the answer appears to be huge, for example if $\sigma^2= 3\times 10^{-3}$ (the value relevant for the formation of PBHs which are decaying today) then
$$ \frac{\beta(\delta_c=1/3)}{\beta(\delta_c=0.45)} \simeq 10^{-6} . $$
However, the change in the value of $\calP_\calR$ required to get a particular value of $\beta$, is only
$$ \frac{\sigma^2(\delta_c=1/3)}{\sigma^2(\delta_c=0.45)}\simeq \frac{0.33^2}{0.45^2} \simeq 0.5 . $$
Authors of PBH papers like to phrase the difference in terms of $\beta$ in order to make the importance of their results exponentially large, but in reality, this normally overstates the importance of the change.

One exception is during the QCD transition when the equation-of-state parameter $\omega$ drops from $1/3$ to $0.25$, about a 25$\%$ reduction, at the time when the horizon mass is one solar mass \cite{Borsanyi:2016ksw}. This reduction in the background pressure makes PBH formation `easier', and the collapse threshold drops from $\delta_c=0.45$ to $\delta_c\simeq0.4$, a reduction of about $10\%$ \cite{Byrnes:2018clq}. This relatively small reduction leads to a 2--3 orders-of-magnitude enhancement in the production of solar mass PBHs compared to the number of PBHs on similar mass scales where $\omega=1/3$ and $\delta_c=0.45$, provided that the amplitude of the primordial power spectrum is sufficiently large and constant over the relevant range of scales. This mass range is of special interest both because they are below the Chandrasekhar mass (about 1.4 solar masses) which is the smallest mass with which a compact object (a neutron star or BH) can form in the late universe through standard astrophysical processes and also because this mass range can be probed by ground-based gravitational wave detectors such as the current LIGO and Virgo instruments. The observation of a sub solar-mass compact object would be a smoking gun signature for a PBH and the detection of even just one such object would have huge implications for our understanding of dark matter and the physics of the very early universe. \\


\section{Observational searches for PBHs}

When studying PBHs as a DM candidate the constraints are normally best phrased in terms of $\fpbh$, but for PBHs which are currently evaporating, or which have already evaporated, it is conventional to use constraints in terms of the initial collapse fraction $\beta$. Recall that neglecting accretion and evaporation the two are related by
\be \fpbh\simeq \frac{\aeq}{\aform} \beta \ee
and this is sometimes used to define $\fpbh$ even in the case that the PBHs have evaporated, even though this means that $\rhopbh=0$ today. Roughly speaking, the observational constraints can be divided into two categories:
\begin{enumerate}
\item Gravitational constraints which more directly relate to $\fpbh$ constraints.
\item Hawking evaporation constraints which are best phrased in terms of $\beta$.
\end{enumerate}

\subsubsection{Microlensing constraints}

If a PBH (or any other sufficiently compact object) passes close to the line of sight to a more distant luminous object, then the light from the source will be gravitationally focused and enhanced. Hence, PBHs can make stars look brighter while they are close to the line of sight between the observer and the star. The strength of the lensing magnification increases with the mass of the object and is largest when the lensing object passes closest to the line of sight to the luminous source. Therefore, the total duration of the luminosity enhancement depends on the mass of the compact object as well as its velocity transverse to the line of sight, with the timescale of the magnification signal varying from a few hours for a $10^{-6} M_\odot$ compact object to a timescale of months for a $10 M_\odot$ mass object.

Historically, searches for compact objects focused on repeating observations of the same patch of the sky every night for days or years. Surveys such as OGLE and EROS were most sensitive to compact objects in the mass range $10^{-6}-1 M_\odot$, and they have found a few lensing events, but not more than would (probably) be expected from compact objects created by standard astrophysical processes such as freely floating planets, and the overall constraint in the mass range they could probe was $\fpbh\lesssim 0.1$. 

Recently, the Hyper-Suprime Cam (HSC) on the Subaru telescope made a very detailed search over just one night for low mass lenses, and this greatly improved the constraints on this mass range, being sensitive enough to provide the constraint of $\fpbh\lesssim 10^{-2}$ for $\Mpbh\sim10^{-9}M_\odot$ and ruling out $\fpbh=1$ for the mass range $10^{-12}M_\odot \lesssim \Mpbh \lesssim 10^{-6}M_\odot$. Originally the HSC collaboration claimed to have constrained even smaller mass compact objects, but this has now been accepted not to be correct because of the finite source effect, which means that once the apparent size in the sky of the lensing object becomes comparable to the apparent size of the object being lensed then gravitational lensing is only effective on part of the surface of the lensed star and not the entire surface. This means that the relative enhancement in the luminosity becomes too small to be detectable \cite{Niikura:2017zjd}. 

\subsubsection{Other gravitational constraints}

For larger masses (more than about a solar mass) there are two other constraints which have been considered for a long time, accretion and DM discreteness effects, plus a much newer gravitational wave constraint.

The accretion of gas onto black holes emits high energy radiation which can be detected, for example there was recently a lot of publicity about the event horizon telescope ``image'' of a supermassive black hole.\footnote{For a review of astrophysical black holes see \cite{Bambi:2019xzp}.} Accretion involves highly non-linear physics and is therefore hard to model robustly, but the constraints from accretion (both at $z=0$ or during recombination when the CMB formed) appear to rule out $\fpbh=1$ for $\Mpbh\gtrsim {\rm few}\, \times M_\odot$ and the constraints become much tighter for larger masses, e.g.~$\fpbh\lesssim10^{-4}$ for $\Mpbh\gtrsim10^2M_\odot$. In general, accretion is not expected to be significant for PBHs with an initial mass much below ten solar masses, and hence light PBHs are expected to have an essentially constant mass (unless they are so small that Hawking evaporation is important) and hence also a constant spin \cite{DeLuca:2020fpg}. 

Discreteness effects (sometimes called dynamical constraints) are caused by very massive PBHs not looking like a `smooth' density field on small scales. For example, dwarf galaxies with masses $10^7-10^9 M_\odot$ could be modified or even destroyed if DM was made out of PBHs with very large masses. In practise these constraints are not as tight as the accretion constraints and hence are not so widely discussed.

The gravitational wave bound is based on the LIGO and Virgo observations of merging compact objects. These detectors are most sensitive in the $1-10^2 M_\odot$ mass range and the tightest constraint on these scales is $\fpbh\lesssim10^{-3}$. Gravitational wave observations provide the most stringent constraint in this mass range. They are also sensitive in the sub-solar mass range but microlensing constraints are stronger on these scales. However, note that these constraints are model dependent since it is not straightforward to estimate the current merger rate of a large population of PBHs. The standard calculation assumes that PBHs form binary pairs in the very early universe, shortly after formation and long before matter-radiation equality, and that many of these binary pairs remain relatively undisturbed until today. Estimating the disruption rate of these `primordial' binaries pairs is a numerically challenging task but there appears to be a consensus that although disruption is a very important effect when $\fpbh\simeq1$ and hence there are many PBHs which can disrupt each other, that for $\fpbh\lesssim10^{-2}$ that disruption becomes relatively rare \cite{Raidal:2018bbj}. Requiring that all of the observed LIGO Virgo merger events were due to primordial black holes requires $\fpbh\simeq 3\times10^{-3}$ (assuming a lognormal mass distribution with a central mass around the $20M_\odot$ but this scenario is strongly disfavoured compared to the alternative (standard) scenario that all of the mergers are due to astrophysical black holes \cite{Hall:2020daa}. A mixed scenario with astrophysical and a subdominant population of primordial black hole mergers works \cite{Hutsi:2020sol,DeLuca:2021wjr}. In practice astrophysical models of black hole formation and mergers are sufficiently uncertain that it would be very hard to prove any specific merger of two compact objects was caused by primordial black holes unless at least one of the component masses is clearly below the Chandrasekhar mass, which is the lowest mass compact object that can form through standard astrophysical processes.

For masses lower than those constrained by the microlensing constraints, there is a relatively wide mass window where $\fpbh=1$ is possible, between $10^{17}-10^{22}$ g or equivalently $10^{-16}-10^{-12} M_\odot$. This window, in which all of the DM could consist of PBHs, is sometimes referred to as the `asteroid mass window'. 

Although the constraints discussed here are based on an unrealistic monochromatic (single) mass function, in practice the constraints do not change by more than a factor of order unity when considering more realistic and broader mass functions \cite{Bellomo:2017zsr}. Hence there is reasonable (but not complete) agreement that even a broad mass function would not allow for all the DM being made out of PBHs with masses far above the `asteroid' mass range. 

\subsubsection{Evaporation constraints and PBH relics}

Black holes are not perfectly black when taking quantum mechanical effects into account, and in fact they radiate energy away with a temperature set by the Hawking radiation from a black hole, which satisfies
$$ T\propto \Mpbh^{-1}. $$
Recalling that the radius of a black hole is proportional to its mass, meaning its surface area is proportional to $\Mpbh^2$, and given that the energy radiated follows a blackbody distribution with total energy proportional to $T^4$ per unit area, the total energy radiated away is proportional to the area times $T^4$, which is proportional to $\Mpbh^{-2}$. Using $E=mc^2$ we can deduce that the rate of mass loss from the black hole satisfies 
$$\frac{d\Mpbh}{d t} \propto \frac{1}{\Mpbh^2} $$
and this can be integrated to find that the lifetime of a BH satisfies
$$t_{\rm evap}\propto \Mpbh^3. $$
Inserting numerical factors one can check that the evaporation timescale equals the age of the universe for a PBH with initial mass $\Mpbh\simeq 10^{15}$ g, and that the Hawking evaporation is non-negligible today if the initial mass satisfies $\Mpbh\lesssim 10^{17}$ g. PBHs which decay today or during the time of recombination when the CMB formed are very tightly constrained by observations, and there are also constraints on the allowed decay of PBHs during BBN. However, the decay of PBHs before BBN begins (corresponding to those with initial mass $\Mpbh\lesssim10^{10}$ g) is extremely hard to constrain using any observations.

When the PBH comes close to completely evaporating the energy being evaporated approaches to the Planck energy at the time the mass becomes comparable to the Planck mass. At this point the semi-classical physics used to derive the Hawking temperature might not remain valid and it remains an open question whether black holes evaporate completely or whether a relic is left behind. If a relic remains it would presumably have approximately a Planckian mass. If Planck mass relics can form, then they could be DM candidates, albeit potentially a `nightmare' DM scenario in which the DM is impossible to detect with any known technology. However, there has recently been a theoretical argument made that relics would gain a large peculiar velocity during the decay process and this could rule them out as a cold dark matter candidate \cite{Kovacik:2021qms}. 

\begin{figure}
\centering{
\includegraphics[width=14 cm,clip=]{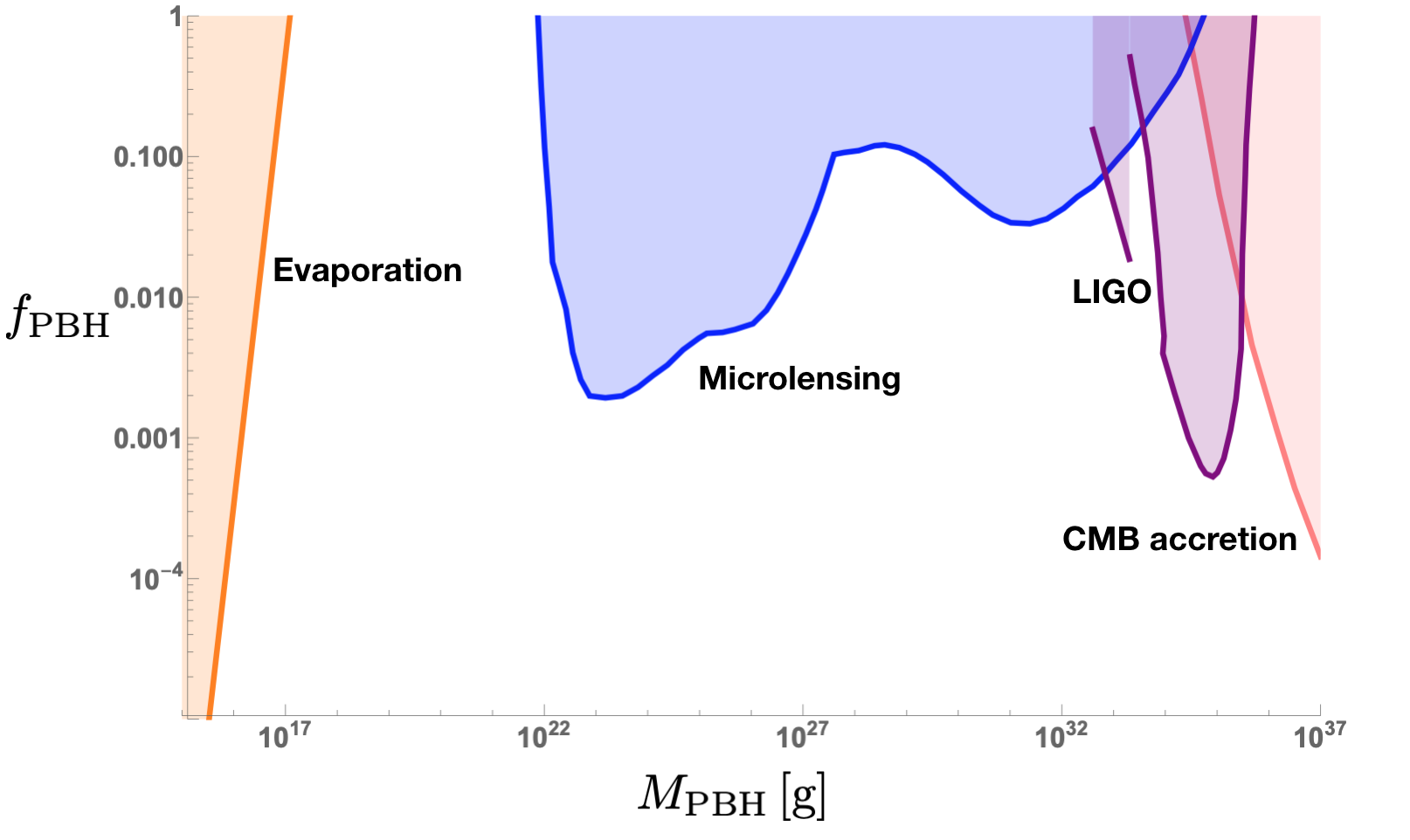}}
\caption{Some of the key constraints on $\fpbh$ showing which types of constraints are most important over a wide range of mass scales. Figure from \cite{Cole:2021cnu}, made with data from \cite{Green:2020jor}.} \label{fig:fPBH}
\end{figure}

Figure \ref{fig:fPBH} shows a summary of the constraints on $\fpbh$ over a wide range of masses. For more details of the observational constraints see \cite{Green:2020jor} and references therein, or for a comprehensive list of constraints phrased in terms of $\beta$ see \cite{Carr:2020gox}.

\subsection{Constraints on the primordial power spectrum}

In this section we will briefly summarise the primary alternative methods to constraining the power spectrum amplitude. On scales larger than about a Mpc, $k\lesssim {\rm Mpc}^{-1}$, observations of the CMB and LSS have provided an accurate measurement of the amplitude. Over a huge range of smaller scales, PBHs provide a weaker constraint as discussed in section \ref{sec:beta}. 

\subsubsection{Spectral distortions}

The early universe before recombination was in thermal equilibrium, and therefore the CMB photons follow a black body distribution. Confirmation of this fact by the COBE FIRAS instrument was a very important piece of evidence in favour of the Big Bang theory. However, the damping of large amplitude density perturbations during certain redshifts/temperatures would act as an energy injection into the baryon-photon plasma, which could lead to a deviation from thermal equilibrium and hence a deviation from a blackbody spectrum \cite{Chluba:2019nxa}. 

The relevant constraint from the non-detection of a cosmic-$\mu$ distortion based on the COBE FIRAS results are roughly $\calP_\calR\lesssim10^{-4}$ in the range $k\sim 1-10^4$ Mpc$^{-1}$, where the smallest constrained scale corresponds to a PBH mass of about $10^4 M_\odot$. Hence, PBHs cannot have a larger mass than this if they were formed by the collapse of large amplitude perturbations shortly after horizon entry, unless the perturbations were strongly non-Gaussian or if there was an early matter dominated epoch taking place while the relevant scales were entering the horizon.

\subsubsection{Gravitational waves} 

As you may have learnt in an introductory cosmology course, at linear order the scalar, vector and tensor perturbations all decouple. On large scales, the perturbations observed via the CMB are so small that we know linear perturbation theory is an excellent approximation. However, on smaller scales where the perturbation amplitude could be much larger, there could be a significant non-linear coupling between different types of perturbations. Of particular interest are the second-order tensor perturbations which are generated by the square of linear scalar perturbations. The full equations showing the coupled evolution of the non-linear perturbations are very complicated, but we can write their order of magnitude schematically as $h^{(2)}\sim \calR^2$, where $h^{(2)}$ denotes the intrinsically second-order tensor (gravitational wave) perturbations. These induce a power spectrum amplitude of the tensor perturbations whose order of magnitude is given by
\be \calP_T\sim \left(h^{(2)}\right)^2\sim \calR^4 \sim\calP_\calR^2. \ee
The corresponding frequency of the waves is given in terms of the scale at horizon entry, $k=aH$, and speed of light, $c$, by
$$ f\simeq c k. $$
For a comprehensive review article about these `induced' gravitational, waves see \cite{Domenech:2021ztg}. Also see \cite{sasakireview} for a review on the many connections between gravitational waves and PBHs.

\begin{figure}
\centering
\includegraphics[width=0.8\textwidth]{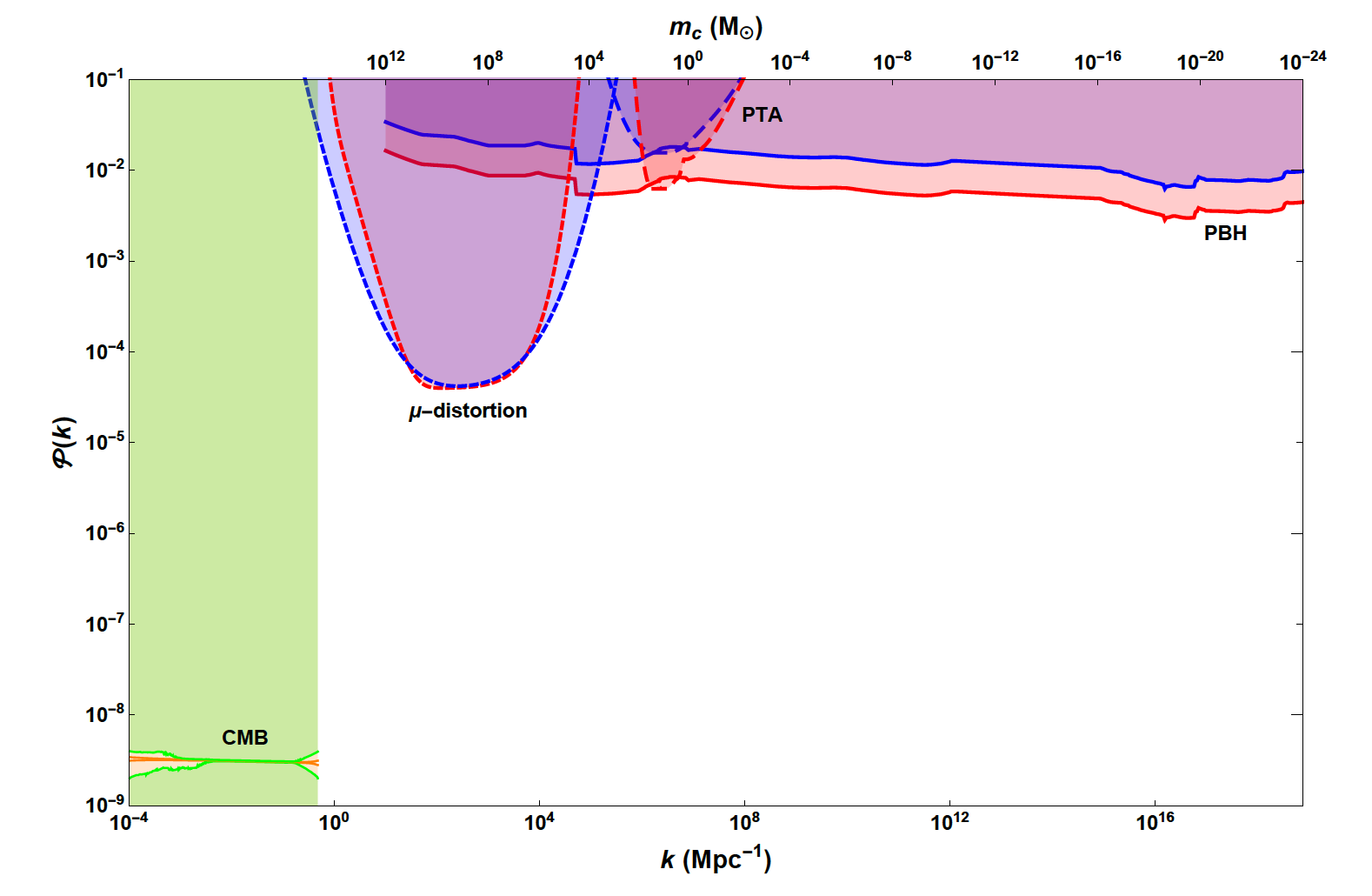} 
\vspace*{-1em}
\includegraphics[width=0.8\textwidth]{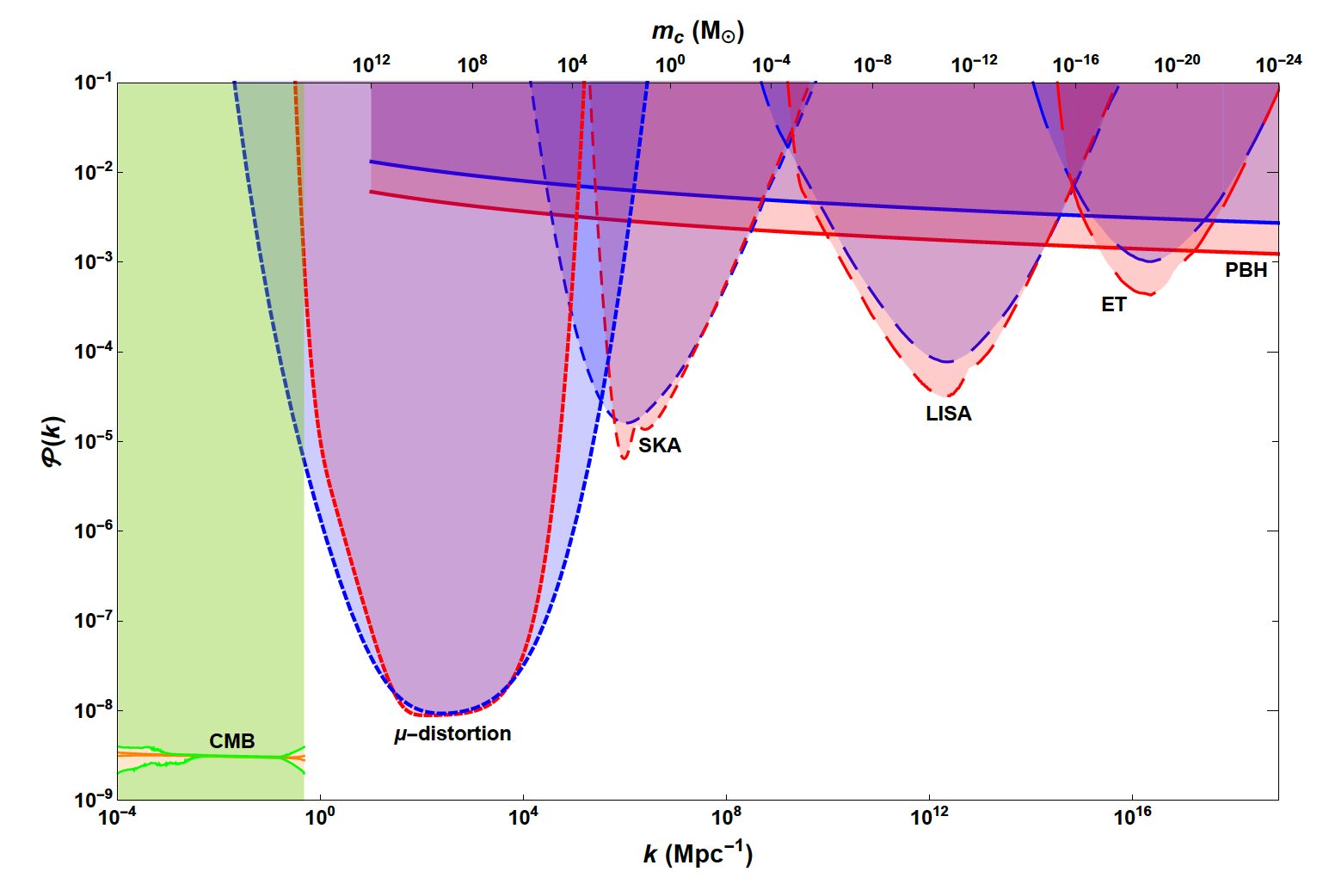}
\caption{Constraints on the amplitude of the primordial power spectrum. On both figures the left hand lines show the amplitude detected by current CMB observations. All other shaded regions show upper bounds on the amplitude, focusing on current constraints in the upper plot and future constraints in the lower plot. For both plots the blue lines (which are weaker for PBH constraints and broader for all other constraints) are based on a narrowly peaked power spectrum, while the red lines show the equivelant constraints for a power spectrum with a broader peak. The PBH constraint lines in the lower plot are the best theoretically possible constraints, based on zero PBHs forming inside the observable universe. The top horizontal axis, $m_c$, shows the central PBH mass corresponding to each value of $k$ shown in the lower axis. For more details see \cite{Gow:2020bzo}. }
\label{fig:PS}
\end{figure}

These second-order tensor perturbations could appear as a stochastic background of gravitational waves. By a nice coincidence the scale corresponding to a horizon mass of 1 solar mass (and the QCD transition) has a frequency in the range which pulsar timing arrays (PTA) can constrain, and the current PTA constraints on the amplitude of the primordial scalar power spectrum are at almost exactly the same amplitude as is required to generate PBHs with this mass. Therefore, if LIGO and Virgo have or do detect any PBHs then we should expect to see a corresponding signal of stochastic gravitational waves. The coincidence of scales and constraints is shown in the upper plot of figure \ref{fig:PS}. We note that the NANOGrav collaboration have recently reported a detection of excess noise in pulsar timing residuals, but they don't have a strong enough signal to determine whether this data is caused by gravitational waves \cite{Arzoumanian:2020vkk}. 

It is also of interest to note that if the DM does consist of asteroid mass PBHs then the LISA space based gravitational wave detector, due to launch in the mid 2030's, is expected to detect an associated signature of stochastic gravitational waves at high significance \cite{Saito:2009jt}. The lower plot of figure \ref{fig:PS} shows this forecasted sensitivity of the LISA instrument, as well as the Einstein Telescope (ET) and the Square Kilometre Array (SKA) constraints on pulsar timing observations. The forecasted constraints on the $\mu$-distortions are based on assuming a PIXIE-like survey \cite{Kogut:2011xw}. 

\subsection{Possible signatures}

Every observational constraint is also a possible signature. However, whilst some signatures could be relatively clearly identified as being due to a PBH, others would be much harder to interpret. For example, a detection of microlensing events could be due to freely floating planets or other astrophysical objects, and it is hard to make a robust estimate of how many such objects could form out of baryonic matter. Likewise, a detection of gamma rays, which could be due to evaporating PBHs, could also be caused by high energy astrophysics or either decaying or annihilating dark matter particles. Nonetheless, there are some tentative hints of PBH signatures in existing data, see \cite{Clesse:2017bsw} for an overview. 

The most promising direct detection signature which is accessible with current instruments would be the discovery for a sub-solar mass compact object \cite{Magee:2018opb}.
If such low mass PBHs are not detected, an alternative direct gravitational wave probe - in the far future with a detector such as the Cosmic Explorer - would be the detection of a very high redshift merger ($z\gtrsim40$), which would be a signature from such early times that stellar objects would not yet have had time to collapse into compact objects \cite{Koushiappas:2017kqm}.

A less direct, but still promising, gravitational wave probe is via the stochastic background of gravitational waves generated at high redshift. The most promising current probe are the PTA searches which form a synergy with the characteristic mass of the LIGO-Virgo detections and the QCD transition, as shown in the upper plot of figure \ref{fig:PS}. This opens the possibility that an analysis of the frequency dependence of the PTA gravitational wave background (if detected and not astrophysical in origin) could be tested against the corresponding mass range of LIGO and other ground-based gravitational wave detectors. 

In the longer term (on a time scale of a few decades) the space-based LISA gravitational wave detector should see a clear signal of a stochastic gravitational wave background if the DM is made out of asteroid mass PBHS. Recalling the logarithmic sensitivity of the scalar power spectrum amplitude to $\fpbh$ (and hence also the associated tensor power spectrum) it becomes clear that the non-detection of a primordial stochastic background by LISA would not only rule out $\fpbh=1$ but even the formation of any PBHs at all in the same mass range.\footnote{Assuming, as is done in these lectures, that the primordial perturbations are close to Gaussian distributed and that PBHs form from the direct collapse of large amplitude density perturbations shortly after horizon entry.} Given that the lower plot of figure \ref{fig:PS} shows the upper bound on the power spectrum amplitude below which zero PBHs would form in today’s observable universe, we can see that in the future there will be almost no remaining mass windows for non-evaporated (or relic) PBHs to have formed, assuming that none of the experiments shown detect the relevant signature used to constrain the power spectrum amplitude. Hence, future searches for PBHs have a bright future, and we can realistically hope to determine whether or not DM is made out of a new particle or primordial black holes. 

\section{Conclusions}

In summary -- whether or not PBHs exist -- the search for them has led to new understandings about the nature of the contents and initial conditions of the universe. Since they require a large amplitude of density perturbations to be generated during inflation, the detection of even just a single PBH would tell us that we need some very special conditions during inflation, namely a phase of ultra-slow-roll at least in the case of single-field inflation, in order to produce them. Furthermore, their detection would also go some way to explaining what makes up the dark matter content of the universe. However, their non-detection is still an incredibly valuable probe, because they are unique in being able to constrain the widest range of scales of the primordial power spectrum due to the fact that they can theoretically form at any mass. Forthcoming searches for PBHs both direct, via for example microlensing or gravitational waves, or indirect, via CMB spectral distortions or the stochastic gravitational wave background, have great promise confirm or deny their existence and potentially settle long-unanswered questions.

\section*{Acknowledgements}
We thank the GGI for the invitation to hold a graduate lecture course on this topic, which formed the basis of these lecture notes. We thank Andreas Mantziris for sending a list of corrections to these lecture notes at the end of the GGI school. CB acknowledges support from the Science and Technology Facilities Council [grant number ST/T000473/1]. PC acknowledges the Institute of Physics at the University of Amsterdam.








\bibliographystyle{JHEP-edit}
\bibliography{bib-file}




\appendix

\section{PBH problems}
\subsection{PBHs from ultra-slow-roll inflation}

For a given model of ultra slow-roll inflation, the power spectrum amplitude at a scale corresponding to 1 solar-mass PBHs is $\mathcal{P}_\mathcal{R}=0.0052$.

\begin{enumerate}[label=(\alph*)]
    \item Assuming a monochromatic population of PBHs, what is the value of the mass fraction $\beta$ at this scale?\\ 
    You may also assume $\sigma^2\sim\mathcal{P}_\mathcal{R}$, note that this is only really valid for a monochromatic power spectrum, which is not actually realistic!
    
    \item What is the corresponding value of $f_{\rm PBH}$?
    
    \item There is a constraint on the power spectrum at this scale, meaning that the amplitude must satisfy $\mathcal{P}_\mathcal{R}<0.004$. How much smaller are the values of $\beta$ and $f_{\rm PBH}$ under this constraint?
    
    \item If the amplitude of the power spectrum depends on the number of e-folds of the  ultra-slow-roll phase $N_{\rm USR}$ as 
    \begin{equation}\label{eq:USR}
        \mathcal{P}_\mathcal{R}\propto \exp(6N_{\rm USR}),
    \end{equation}
    how many fewer e-folds of ultra-slow-roll are required so as to adhere to this constraint?
    
\end{enumerate}

Use $\delta_c=0.45$ 
for this question.
\subsection{PBHs forming during matter domination}
\begin{enumerate}[label=(\alph*)]
    \item The power spectrum is not actually directly proportional to $\exp(6N_{USR})$ as in equation \eqref{eq:USR}. Other than the amplitude of the peak, how else will the spectrum change if the number of e-folds of ultra-slow-roll inflation changes?
        
    \item If there is an early-matter dominated phase, then the relationship between primordial black hole abundance and the power spectrum is
    \begin{equation}\label{eq:matter}
        \beta_0\simeq0.056\sigma^5
    \end{equation}
    where $\beta_0=(t_1/t_i)^\frac{2}{3}\beta$, $t_i$ is the time of formation, and $t_1$ is the time that the early-matter dominated phase ends.
    
    If PBHs of $10^{20}\,{\rm g}$ make up all of the dark matter (still possible!) then what is the required amplitude of the power spectrum, if there was an early matter dominated phase that ended at $t=10^{-14}\,{\rm s}$?
    \end{enumerate}

\section{PBH solutions}
\subsection{PBHs from ultra-slow-roll inflation}

\begin{enumerate}[label=(\alph*)]
    \item The exact answer will depend on whether the approximation for $\beta$ is used. For the expression that uses the complementary error function:
    \begin{equation}
        \beta={\rm erfc}\left(\frac{\delta_c}{\sqrt{2}\sigma}\right)=4.365\times10^{-10}.
    \end{equation}
    
    \item Using the following expression, we find $f_{\rm PBH}\approx1$: \begin{equation}
        f_{\rm PBH}=\left(\frac{M_{PBH}}{M_{eq}}\right)^{-\frac{1}{2}}\frac{\beta}{\Omega_{\rm DM}}\approx1.
    \end{equation}
    \item For example, $\beta=1.1\times10^{-12}$ and  $f_{PBH}=0.0025$. Your answer for $f_{PBH}$ should be around 3 orders of magnitude smaller than part (a).
    
    \item Using the following expression: \begin{equation}
        \frac{\mathcal{P}_1}{\mathcal{P}_2}=\exp(6(N_1-N_2)),
    \end{equation}
    and plugging the numbers in from above we find $N_1-N_2=0.044$.
\end{enumerate}

\subsection{PBHs forming during matter domination}
\begin{enumerate}[label=(\alph*)]
    \item 
    If the start of the USR phase is kept fixed, changing the number of e-folds of USR will also change the position of the peak. This means that you can't just tune the number of PBHs of a given mass that a specific inflationary model and resulting power spectrum produces by altering the number of USR e-folds alone - you'd then also need to change the start point of USR, which corresponds to the field value of the feature in the potential.
    
    \item 
    First, calculate the time at which PBHs with mass $10^{20}{\rm g}$ are formed:
    \begin{equation}
        t_i=\frac{4GM_{PBH}}{3c^3}=3.3\times10^{-19}{\rm s}.
    \end{equation}
    Then for $f_{PBH}=1$, the standard PBH mass fraction, is $\beta=9.8\times10^{-17}$. This means that $\beta_0$, which is the `matter-domination mass fraction' which accounts for the fact that $\Omega_{PBH}$ remains constant during a matter dominated phase, is 
    \begin{equation}
        \beta_0=\left(\frac{10^{-14}}{3.3\times10^{-19}}\right)^\frac{2}{3}\times9.8\times10^{-17}=9.55\times10^{-14}.
    \end{equation}
    Putting this into equation \eqref{eq:matter} and again using $\mathcal{P}\sim\sigma^2$, we find
    \begin{equation}
        \mathcal{P}_\mathcal{R}=1.96\times10^{-5},
    \end{equation}
    which is much smaller than any amplitude which would produce even a negligible fraction of the dark matter if they formed during radiation domination.
\end{enumerate}

\end{document}